\documentstyle[preprint,aps]{revtex}

\draft

\begin{document}

\title{From semiclassical transport to quantum Hall effect under low-field Landau quantization}
\author{ D. R. Hang$^{a,b,*}$, C. F. Huang$^{c}$, Y. W. Zhang$^{d}$, H. D. Yeh$^{c}$, J. C. Hsiao$^{c}$, and H. L. Pang$^{c}$}

\address{$^{(a)}$ Department of Materials Science and Optoelectronic Engineering, National Sun Yat-sen University, Kaohsiung 804, Taiwan, R.O.C. \newline
$^{(b)}$ Center for Nanoscience and Nanotechnology, National Sun Yat-sen University, Kaohsiung 804, Taiwan, R.O.C. \newline
$^{(c)}$ National Measurement Laboratory, Center for Measurement Standards, Industrial Technology Research Institute, Hsinchu 300, Taiwan, R.O.C. \newline
$^{(d)}$ Institute of Materials Science and Engineering, National Sun Yat-sen University, Kaohsiung 804, Taiwan, R.O.C.
\newline $^{*}$ e-mail: drhang@mail.nsysu.edu.tw}

\date{\today}

\maketitle

\begin{abstract}
The crossover from the semiclassical transport to quantum Hall effect is studied by examining a two-dimensional electron system in an AlGaAs/GaAs heterostructure. By probing the magneto-oscillations, it is shown that the semiclassical Shubnikov-de Haas (SdH) formulation can be valid even when the minima of the longitudinal resistivity approach zero. The extension of the applicable range of the SdH theory could be due to the damping effects resulting from disorder and temperature. Moreover, we observed plateau-plateau transition like behavior with such an extension. From our study, it is important to include the positive magnetoresistance to refine the SdH theory.
\end{abstract}
\newpage

\newpage


Considerable efforts have been made to understand how Landau quantization affects the magneto-transport properties of two-dimensional electron systems (2DESs) under a perpendicular magnetic field $B$. It is well-known that Landau quantization can modulate the density of states and induce magneto-oscillations, which are periodic with respect to the inverse of $B$. The 2D Shubnikov-de Haas (SdH) theory, derived from a semiclassical approach, acts as the conventional tool to describe the low-field oscillations [1-5]. In practice, the analysis of low-field oscillations provides a common way to obtain three basic parameters of a 2DES, the carrier concentration, the quantum mobility, and the effective mass. In contrast, to explain the integer quantum Hall effect (IQHE) appearing at higher $B$, we shall consider quantum localization effects [1,6-9]. In the IQHE regime, there are a series of quantum Hall states characterized by quantized Hall plateaus and zero longitudinal resistivity. The magnetic-field-induced phase transitions in the IQHE provide good examples of continuous quantum phase transitions [10,11]. Universal properties based on the scaling theory and the modular symmetry have been investigated in the phase transitions in the IQHE [12-15]. While the universalities can be broken because of some unexpected factors, it has been shown that features of the scaling theory and modular symmetry can still be found after suitable analysis [16]. 

It should be noted that the quantum localization effects are also important even as $B$ approaches zero in the standard theory of the IQHE [7]. The low-field insulator induced by such effects has been observed in 2DESs with large disorder [17,18,19]. For a typical 2DES, in reality, the localization length becomes much larger than the realistic sample size with decreasing $B$ [7,17]. In this way the semiclassical SdH theory, in which the localization effects are ignored, remains valid at low $B$ for most 2DESs while quantum localization effects are still important to the appearance of the IQHE at high $B$ [1]. Therefore, as the magnetic field $B$ is increased, the crossovers from classical (semiclassical) transport to IQHE are expected in the low-field Landau quantization for a wide range of disorder. 

Although successful theories have been developed to understand the magneto-transport properties of 2DESs, there still exist many unresolved questions and ambiguities. At high $B$, an experimental study inconsistent with the scaling theory has been reported [20]. On the other hand, at low $B$ there are also debates on the validity of the Lifshitz-Kosevich formula, originally derived for the three-dimensional Fermi liquid, to be applied in the 2D SdH theory [4,5,21]. It has been suggested theoretically in Ref. 4 that the damping due to disorder and temperature is important to transfer such a formula to 2D cases. Besides, while band parameter like effective mass derived from magneto-oscillations is conventionally taken as a constant, the field dependence has been studied by several groups [22-24]. Our group has also reported the enhancement of the effective mass of a 2D GaN electron system with increasing $B$. [22] Therefore more experimental investigations are necessary to probe the SdH theory. To further understand the exact behavior in the crossover from semiclassical to IQHE regime, we look into the quantum magneto-oscillations and low-field IQHE in a 2DES in an AlGaAs/GaAs heterostructure. The effective mass $m^{*}$ is a well-established constant in a 2D GaAs electron system when carrier concentration lies within the typical range. Therefore, we can probe the applicable range of SdH formulation in such a system by examining the value of $m^{*}$ while for some other materials it is suitable to investigate the meaning of band parameters by using this formulation. We show that such a semiclassical formulation can be valid even when the minima of the longitudinal resistivity approach zero. In addition, the positive magnetoresistance should be taken into account to refine the SdH formulation. 

The sample used for our study is an AlGaAs/GaAs heterostructure, in which a 2DES resides in the GaAs side of the heterojunction. The 2D channel was followed by a 15 nm spacer layer of AlGaAs, a 40 nm layer of AlGaAs doped with Si at $1 \times 10 ^{18} cm ^{-3}$ and a 12 nm GaAs cap layer doped at $1 \times 10 ^{18} cm ^{-3}$. A Hall pattern was made by the standard lithography and etching process. Magnetotransport measurements were done with a 12 Tesla superconducting magnet and a He$^{4}$ refrigerator. Figure 1 shows the four-terminal longitudinal and Hall magnetoresistivity $\rho _{xx}$ and $\rho _{xy}$  at temperature $T = 4.2$ K. When a small perpendicular magnetic field is applied, the 2DES is governed by the classical transport theory, so we observe $\rho _{xx} \sim$ constant and $\rho _{xy} = B/ne$. Here $n$ is the carrier concentration and $e$ is the electronic charge. With gradually increasing magnetic fields, the 2DES enters the semiclassical regime. A set of magneto-oscillations, commonly known as Shubnikov-de Haas oscillations whose periodicity is determined by the 2D carrier density, can be observed. The 2D carrier concentration can be obtained from the low-field oscillations to be $3.17 \times 10 ^{11} cm ^{-2}$. The classical mobility $\mu _{c}$ is estimated to be $5.3 \times 10 ^{5} cm ^{2} /V$-$s$. from $\rho _{xx} (B = 0) = 1/ne \mu _{c}$. As shown in Fig. 1, at higher magnetic fields, in which the 2DES is in the strong localization regime, we can observe well-developed quantum Hall states with $\rho _{xx} \rightarrow 0$ and $\rho _{xy} = h / \nu e ^{2}$ and of filling factors down to $\nu = 2$.

The low-field oscillation amplitude $\Delta \rho _{xx}$ at finite temperatures obtained from semiclassical SdH theory is given by [2]
\begin{eqnarray}
\Delta \rho _{xx} (B,T) = 4 \rho _{0} D ( m ^{*} , T ) exp (-\frac{\pi}{\mu _{q} B} ) 
\end{eqnarray}
where $\rho _{0}$ is a constant, $\mu _{q}$ is the quantum mobility, the temperature factor $\chi/ sinh \chi$, $\chi = 2 \pi ^{2} k _{B} T / \hbar e B$, $\hbar$ is the reduced Planck constant, $k _{B}$ is the Boltzmann constant, and $m ^{*}$ is the electron effective mass. This equation is expected to hold true for small magneto-oscillations before well-developed quantum Hall states and zero longitudinal resistivity appear with increasing $B$. In addition, the constant $\rho _{0}$ is expected to be the zero-field longitudinal resistivity $\rho _{xx}(B = 0)$ although there are reports on the deviations [2].

The detailed temperature dependence of low-field magneto-oscillations in $\rho _{xx}$ is shown in Fig. 2. With increasing temperature, the amplitude of the oscillations is damped. At relatively high temperatures where $\chi > 1$, Eq. (1) can be further simplified to yield:
\begin{eqnarray}
ln \frac{\Delta \rho _{xx} (B,T)}{T}= C _{1} - \frac{2 \pi ^{2} k _{B} m ^{*}}{ \hbar e B} T
\end{eqnarray}
where $C _{1}$ is a constant. This equation provides a powerful way to obtain the carrier effective mass in magnetotransport measurements according to the semiclassical SdH theory.   

As $\nu < 9$, as shown in the inset of Fig. 2, the spin-splitting is resolved and the enhancement of exchange spin gaps should be considered [25]. To focus on the range of the semiclassical transport theory, we have analyzed the oscillating amplitudes of filling factors from 29 to 9, where the spin enhancement effect can be ignored. Figure 3 (a) shows the fitting for the effective mass using Eq. (2) for two filling factors, 29 and 9, which correspond to the boundary of the analyzed filling-factor region. At $\nu = 29$, where the oscillation amplitude is reasonably small to follow the SdH theory, an effective mass value of $0.069 m _{0}$ is found, which is in good agreement with the expected value $0.067 m _{0}$. When the filling factor lowers as a consequence of increasing the magnetic field, the minima of $\rho _{xx}$ approach zero as $T$ decreases for $\nu < 14$, where quantized plateaus can be observed in $\rho _{xy}$. Therefore for $\nu < 14$ the theory for IQHE should be considered since it is expected to go beyond the scope of SdH formulation. However, we can see in Fig. 3 (a) that Eq. (2) is still valid at $\nu = 9$ with $m ^{*} = 0.0693 m _{0}$, which is close to $0.067 m _{0}$. To see whether it is just a coincidence, in Fig. 3 (b) we show the complete result of the effective mass value based on Eq. (2). It is found that the obtained effective masses throughout the region investigated possess a striking consistency with an averaged value of $0.0688 m _{0}$. The effective mass value fluctuates within an extent of only $0.8 \%$ even when the IQHE appears at $8 < \nu < 14$. The obtained result indicates that the precision of effective mass obtained by the semiclassical formula remains good even as minima of $\rho _{xx} \rightarrow 0$ and $\rho _{xy} = h / \nu e ^{2}$.  

From Eq. (1), we have
\begin{eqnarray}
ln[\frac{\Delta \rho _{xx} }{ \rho _{xx} (B=0) D (m ^{*}, T)}] = C _{2} - \frac{\pi}{\mu _{q}} \frac{1}{B}
\end{eqnarray}
where $C _{2}$ is a constant. The inset of Fig. 4 shows the plot of $ln[ \Delta \rho _{xx} / \rho _{xx} (B = 0) D( m ^{*}, T)]$ vs. $1 / B$, obtained with the averaged effective mass value. We obtain a quantum mobility $\mu _{q} \sim 5.38 \times 10 ^{4} cm ^{2}/V$-$s$ from the slope of the graph and the constant $C _{2}$ equals 1.55, which is close to the expected value ln4.

To further examine the SdH theory, we proceed to rearrange Eq. (1) as
\begin{eqnarray}
\frac{\Delta \rho _{xx} (B) }{ 4 \rho _{0} exp (- \pi / \mu _{q} B) } = D (m ^{*},T) 
\end{eqnarray}
In this form, the relation can be easily checked by plotting $[\Delta \rho _{xx} / 4 \rho _{0} exp (- \pi / \mu _{q} B ) ]$  with respect to $T / B$, as shown in Fig. 4. If Eq. (1) is a valid description, according to Eq. (4), the amplitudes taken with different $T / B$ should collapse on the solid line given by calculated $D (m ^{*}, T)$ with the obtained parameters. It is found that the entire data points coincide with the theoretical predictions. Even for $8 < \nu < 14$, in which minima of $\rho _{xx}$ approach zero with decreasing $T$, as shown by solid symbols, the temperature factor is still followed well. The agreement with Eq. (4) throughout the region investigated indicates that the semiclassical SdH formula remains a good description in the low-field IQHE in our study.

In the conventional SdH theory, Eq. (1) is derived for low-field situations where the oscillations of the density of states $\Delta g$ are much smaller than the zero-field density of states [6]. As the minima of $\rho _{xx}$ approach zero, $\Delta g$ is no longer small and the quantum diffusion model is considered for high-mobility samples at low temperatures where Eq. (1) is invalid [6,26]. However, the applicable range of semiclassical SdH theory could expand under the damping effects due to temperature and disorder [4]. It is shown that such effects play important roles in transferring three-dimensional Lifshitz-Kosevich formula to the 2D SdH theory [4]. In fact, different mechanisms are responsible under different types of disorder [27]. We note that the mobility of our sample is lower than that in Ref. 26, and the experiments are performed at high temperatures where $\chi > 1$. Our study reveals the extension of semiclassical SdH theory to the IQHE under strong damping effects.  

In the conventional SdH theory, we have 
\begin{eqnarray}
\rho _{xx} \sim \rho _{xx} (B=0) + \Delta \rho _{xx}
\end{eqnarray}
as higher order terms are ignored. Because $\rho _{xx} \geq 0$, the violation of SdH formula can be expected when the oscillation term $\Delta \rho _{xx} > \rho _{xx} (B = 0)$. However, in our study, Eq. (1) holds true even when $\Delta \rho _{xx}$ becomes larger than $\rho _{xx} (B = 0)$ with decreasing $T$ as $\nu < 16$. At $T =1.9$ K, as indicated by the dashed line in the inset of Fig. 2, we have an apparent positive magnetoresistance as the nonoscillatory background after averaging the magneto-oscillations. Because of such a nonoscillatory part, Eq. (1) remains true when $\Delta \rho _{xx} > \rho _{xx} (B = 0)$ without inducing the negative resistivity. Different mechanisms have been introduced for the positive magnetoresistance, [28-30] and our study shows its importance to extend the applicable range of the SdH formula.  

To further study the crossover from the semiclassical transport regime to IQHE, we also investigate the behaviors of $\rho _{xy}$ when the semiclassical SdH formula Eq. (1) is valid. Figure 5 shows the temperature dependence of $\rho _{xy}$ at $B \sim 1$ Tesla, near which the positive magneto-resistance is important to the extension of SdH formula for $\Delta \rho _{xx}$ as mentioned above. In addition to Hall plateaus, there is a temperature-independent point at $B = 1.022$ Tesla between the plateaus of the high filling factor $\nu = 14$ and 12. Similar temperature-independent points exist between the plateaus $\nu = 12$ and 10, and $\nu = 10$ and 8 as well. Temperature-independent points are expected in both $\rho _{xx}$ and $\rho _{xy}$ at critical magnetic fields of plateau transitions under high-field quantum localization, which is important to the standard theories for IQHE [7,31]. In our study, however, such points appear only in $\rho _{xy}$. It has been reported that quantum localization leading to IQHE is more robust in $\rho _{xy}$ than in $\rho _{xx}$ [1]. By inverting the corresponding conductivities, [13,32] the horizontal dash line in Fig. 5 indicates the expected $T$-independent position, which deviates a little from the experimental one. We note that the deviation may exist even under the high-field localization [7,13]. Alternatively, the features of IQHE in $\rho _{xy}$ can be explained by fixing the chemical potential without considering quantum localization [33,34]. Therefore more studies are necessary to clarify the origins of $T$-independent points between adjacent Hall plateaus when $\Delta \rho _{xx}$ follows SdH formula. 

In conclusion, we report magnetotransport measurement on the 2DES in an AlGaAs/GaAs heterostructure to study the crossover from semiclassical transport to strong localization. Both the longitudinal and Hall resistivities are investigated in such a crossover. While fixed points appear in $\rho _{xy}$ with increasing $B$, we found that semiclassical SdH formula is still valid for the magneto-oscillations in $\rho _{xx}$. Such a formula, in fact, survives even when the minima of $\rho _{xx}$ approach zero at low temperature. The extension of applicable range of the SdH theory could be due to the damping effects resulting from disorder and temperature. It is shown that we should incorporate the positive magnetoresistance to refine the SdH formula. It is suggested that more studies are required to explain the coexistence of plateau-plateau transition-like behavior and the semiclassical SdH formula. 

This work is supported by the National Science Council of the Republic of China under grant no: NSC 94-2112-M-110-009. D. R. Hang acknowledges financial support from ACORC and Aim for the Top University Plan of National Sun Yat-sen University, Taiwan.

\centerline{Figure Captions}
Figure 1: The longitudinal and Hall resisitivity at the temperature 4.2 K. We can observe Shubnikov-de Haas oscillations at lower magnetic fields and quantum Hall plateaus at higher magnetic fields. \newline
Figure 2: The temperature dependence of longitudinal resistivity. The amplitudes of the oscillations are damped when the temperature is increased. Analysis of the amplitudes of quantum magneto-oscillations is done between $\nu$ = 29 and $\nu$ = 9, as is indicated by the arrows. The dashed line in the inset shows the nonoscillatory background at $T$ =1.9 K obtained by averaging the magneto-oscillations. \newline
Figure 3: (a) According to the semiclassical theory, the effective mass can be extracted by plotting $ln( \Delta \rho _{xx} / T )$ versus $T$. (b) The obtained values are close to the expected value $0.067 m _{0}$ not only in the initial small SdH oscillations, but also in the low-field IQHE where fixed points in Hall resistivities are observed. \newline 
Figure 4: The inset shows $ln[\Delta \rho _{xx} / \rho _{xx} (B = 0) D(m ^{*}, T)]$ as a function of inverse magnetic field, from which the quantum mobility can be obtained. The plot of $\Delta \rho _{xx} / 4 \rho _{0} exp (- \pi / \mu _{q} B)$ with respect to $T / B$ can be done accordingly for various fixed temperatures. The symbols squares, circles, up triangles, down triangles and diamonds are for the points at $T$ = 1.9 K, 2.6 K, 3.2 K, 3.7 K and 4.2 K, respectively. The solid symbols for each temperature stand for conditions as $B$ > 1 Tesla where minima of $\rho _{xx}$ approach zero. The numerical evaluation of $D (m ^{*}, T) = \chi / sinh \chi$ as a function of $(T / B)$ is shown as the solid line. \newline
Figure 5: Between the plateaus of the high filling factor $\nu$ = 14 and 12, we observed the temperature-independent point in $\rho _{xy}$, as shown by the arrow. Such a point is close to the expected universal value indicated by the horizontal dash line. 

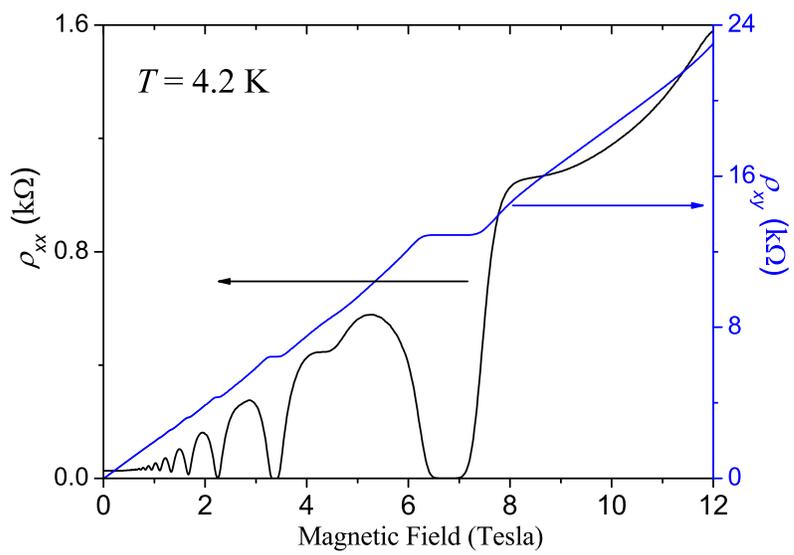
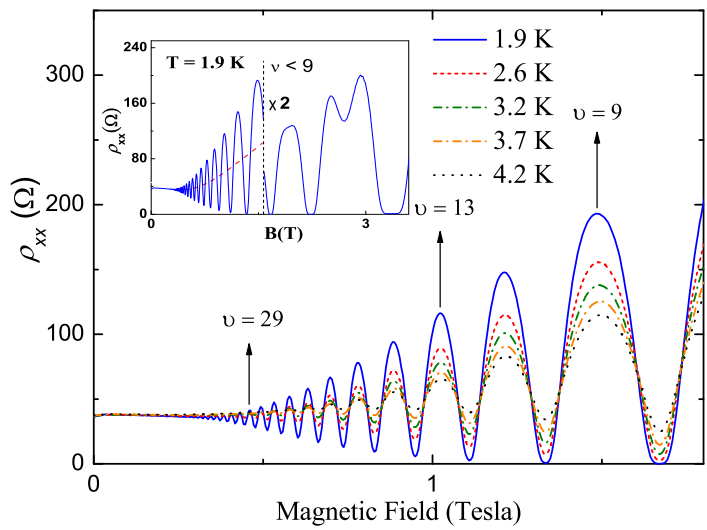
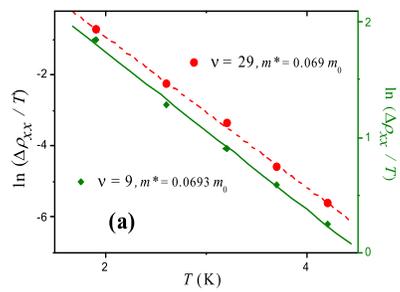
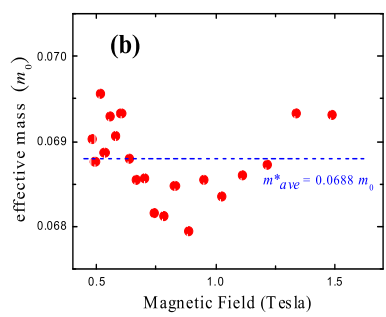
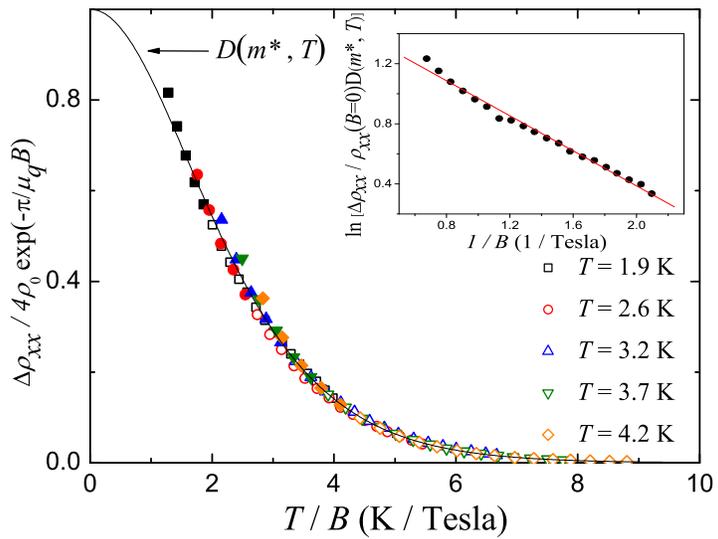
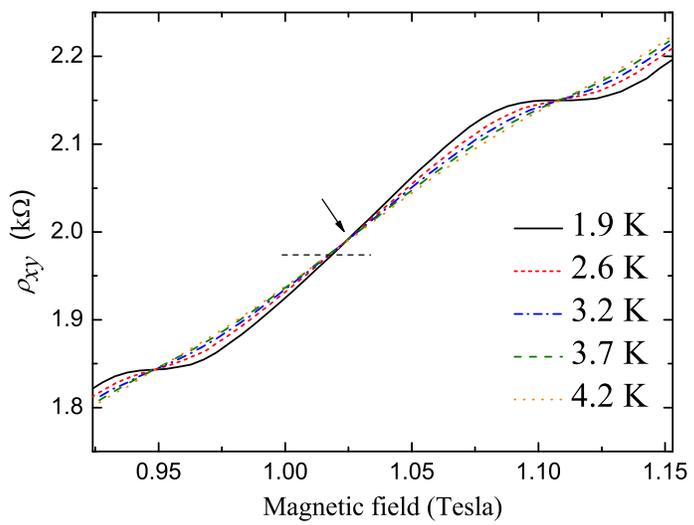
\end{document}